\documentclass[fleqn,10pt]{wlscirep}
\usepackage[utf8]{inputenc}
\usepackage[T1]{fontenc}
\usepackage{times,color,amsfonts,amssymb,amsbsy,amsthm,amsmath,graphics,graphicx,hyperref,bm,bbm,makecell,mathtools,dsfont,upgreek,fancyhdr,multirow}
\usepackage{tabularx}
\usepackage{appendix}
\usepackage{filecontents}

\title{Benchmarking of Quantum Protocols}

\author[1,+]{Chin-Te Liao}
\author[2,4,*,+]{Sima Bahrani}
\author[5,6]{Francisco Ferreira da Silva}
\author[1,2,3]{Elham Kashefi}
\affil[1]{VeriQloud, Paris, France}
\affil[2]{School of Informatics, University of Edinburgh, Edinburgh, United Kingdom}
\affil[3]{CNRS, LIP6, Sorbonne Universit{\'e}, Paris, France}
\affil[4]{High Performance Networks Group, Merchant Venturers Building, University of Bristol, United Kingdom}
\affil[5]{QuTech, Delft University of Technology, Lorentzweg 1, 2628 CJ Delft, The Netherlands}
\affil[6]{Kavli Institute of Nanoscience, Delft University of Technology, Lorentzweg 1, 2628 CJ Delft, The Netherlands}

\affil[*]{si.bahrani@gmail.com}
\affil[+]{these authors contributed equally to this work}

\begin{abstract}
Quantum network protocols offer new functionalities such as enhanced security to communication and computational systems. Despite the rapid progress in quantum hardware, it has not yet reached a level of maturity that enables execution of many quantum protocols in practical settings. To develop quantum protocols in real world, it is necessary to examine their performance considering the imperfections in their practical implementation using simulation platforms. In this paper, we consider several quantum protocols that enable promising functionalities and services in near-future quantum networks. The protocols are chosen from both areas of quantum communication and quantum computation as follows: quantum money, W-state based anonymous transmission, verifiable blind quantum computation, and quantum digital signature. We use NetSquid simulation platform to evaluate the effect of various sources of noise on the performance of these protocols, considering different figures of merit. {We find that to enable quantum money protocol, the decoherence time constant of the quantum memory must be at least three times the storage time of qubits. Furthermore, our simulation results for the w-state based anonymous transmission protocol show that to achieve an average fidelity above 0.8 in this protocol, the storage time of sender's and receiver's particles in the quantum memory must be less than half of the decoherence time constant of the quantum memory. We have also investigated the effect of gate imperfections on the performance of verifiable blind quantum computation. We find that with our chosen parameters, if the depolarizing probability of quantum gates is equal to or greater than 0.05, the security of the protocol cannot be guaranteed. Lastly, our simulation results for quantum digital signature protocol show that channel loss has a significant effect on the probability of repudiation.}
\end{abstract}
\begin{document}
\flushbottom
\maketitle

\section*{Introduction}
In recent years, quantum technologies have seen significant advancements \cite{acin2018quantum,wehner2018quantum,pirandola2020advances,pirandola2015advances}. The rapid development in quantum hardware components such as single-photon detectors and quantum memories promises a vision of small-scale and large-scale quantum networks with real world applications \cite{wallucks2020quantum,wang2019efficient,korzh2020demonstration}. Quantum networks offer new functionalities and services that are not possible in their classical counterpart. {Prominent examples are secure communication and computation enabled by quantum cryptography, quantum secure direct communication (QSDC), blind quantum computing and distributed secure quantum machine learning \cite{caleffi2020rise,pirandola2020advances,cuomo2020towards,sheng2021one,chen2018three,long2021drastic,qi202115,zhou2020device,sheng2017distributed}.} Whereas quantum networks are not meant to replace existing classical ones, they have a great potential to extend the capabilities of classical networks. 

Quantum protocols, as the use cases of quantum networks, offer unique communication and computation features. The most well-known example is quantum key distribution (QKD), which provides forward secrecy. Moreover, quantum cryptographic protocols such as quantum money and quantum digital signature (QDS) enable guaranteeing unforgeability with desired security level \cite{bozzio2018experimental,amiri2017quantum,kumar2019practically,amiri2016secure,wallden2015quantum}. Another significant example is verifiable blind quantum computation (VBQC), which enables delegated quantum computation while preserving privacy \cite{broadbent2009universal,fitzsimons2017unconditionally}. Whereas the advantages offered by these protocols are promising, their commercial deployment requires several steps to be taken.

One major requirement in the development of quantum protocols in real world is the evaluation of their performance in various aspects such as security, required resources, and scalability. Such detailed performance analysis is required to include different sources of imperfection in the practical implementation. This type of benchmarking is mainly important due to the gap between the analyses provided by the academia community in the proposed theoretical protocols and the requirements recommended by experimentalists. Furthermore, it will provide us a tool to compare different quantum protocols with the same functionality proposed in the literature, e.g., different quantum token protocols. Such comparison is crucial in determining the commercial applications and use cases of quantum protocols. 

Another prerequisite for commercial deployment of quantum protocols is to benchmark them against classical and post-quantum protocols with the same functionality. For instance, it would be desirable to compare QKD to symmetric key encryption methods in terms of security, required resources, scalability, and forward secrecy. Another significant example is quantum secure multi-party computation (SMPC) \cite{ben2006secure}, and its benchmarking against classical SMPC. While this type of benchmarking is of paramount importance, it requires the data provided by the evaluation of quantum protocols in practical settings. 

The two types of benchmarking mentioned above, are crucial prerequisites for designing quantum networks. Quantum networks require classical communication for various purposes such as synchronization and control messages. Besides, some quantum protocols, e.g., anonymous transmission, consist of intertwined quantum and classical sub-algorithms. Such integration of quantum and classical building blocks requires a detailed analysis of the interaction between them. In particular, any error or delay in classical communication may adversely affect the performance of quantum sub-algorithms. For instance, the delay in classical messages may increase the decoherence of quantum states in quantum memories. Therefore, to design quantum networks efficiently, it is necessary to determine the impact of classical messages/sub-algorithms involved in the quantum protocols. 

One approach to benchmark quantum protocols is to investigate their performance considering fixed values for system parameters and desired figures of merit. This method will help us to evaluate the effect of specific protocol/hardware parameters and examine the feasibility of practical implementation considering currently achievable parameter values. Another method for benchmarking of quantum protocols is to consider target values for our desired figures of merit and determine the minimum requirements at the hardware level to achieve them. This method, previously proposed in \cite{da2021optimizing}, enables us to optimize system parameters and determine minimum viable requirements to achieve specific target values for figures of merit. We refer to this method as {\it{backward benchmarking}}. In this paper, we mainly focus on the first method of benchmarking. Nevertheless, we provide an example of backward benchmarking by adapting this method for quantum money protocol. 

NetSquid is a software tool which provides a platform for simulating quantum networks and quantum computing systems \cite{coopmans2021netsquid}. This software enables simulation of quantum networks considering various aspects such as physical layer characteristics and control plane. The design of NetSquid is based on discrete-event simulation, which provides us a powerful tool to simulate the decoherence of quantum states by time and analyse the noise in quantum systems accurately. For instance, time-dependent noise affecting the quantum states stored in a quantum memory can be simulated effectively. 

In this paper, we use NetSquid to simulate several quantum protocols and analyse their performance in the presence of various imperfections in the system. In the following sections, benchmarking of quantum money protocol, W-state based anonymous transmission, VBQC, and QDS are presented. Moreover, we provide an example of the backward benchmarking by applying this method to quantum money protocol. 
\section*{Benchmarking of quantum money protocol}
\label{Sec_Qtoken}
Private-key quantum money with classical verification enables a trusted bank to provide unforgeable banknotes to clients. Here, we consider the quantum money protocol proposed in \cite{bozzio2018experimental}. The steps of the protocol are as follows:

1) Bank randomly chooses $n$ qubit pairs from the following set:
\begin{eqnarray}
S_{pair}=\{\left |0+\right \rangle, \left |0-\right \rangle\,\left |1+\right \rangle,\left |1-\right \rangle,\left |+0\right \rangle,\left |-0\right \rangle, 
\left |+1\right \rangle,\left |-1\right \rangle\}\quad\quad\quad\quad\quad\quad\quad\quad\quad\quad,
\end{eqnarray}
and sends them as banknote to the client. 

2) Client stores the received qubits in quantum memory. 

3) Client waits for $T$ seconds. Then, she allows the verifier  to access the banknote. 

4) The verifier randomly chooses between the two bases X and Z, and measures all $2n$ qubits in the chosen basis. 

5) Bank and verifier communicate via a classical channel and check the measurement outcomes. If the number of valid outcomes from the qubits in the chosen basis is larger than a predetermined threshold, bank verifies the validity of the banknote. 

In a practical implementation of this protocol, various sources of loss and noise adversely affect the performance of the system. Table~\ref{noise_loss_qtoken} shows major sources of loss and noise for this protocol. 
\begin{table*}[t]
	\caption{Major sources of noise and loss in quantum money protocol.}
	\centering 
	\begin{tabular}{|c |c|} 
		\hline
		\hline
		Protocol Step & Major sources of noise and loss \\
		\hline 
		\hline
		\multirow{2}{*}{Banknote preparation and transmission to user} & decoherence \\
		\cline{2-2} 
		&  transmission loss\\
		\hline
		Storage in quantum memory & noise introduced by quantum memory\\
		\hline
	\multirow{2}{*}{measurement} & loss introduced by measurement \\ 
		\cline{2-2}
		& measurement error \\
		\hline
		\end{tabular}
	\label{noise_loss_qtoken}
\end{table*} 

\subsection*{Figures of merit}
\subsubsection*{Probability of correctness}
One of the main figures of merit for this protocol is the probability of successful verification assuming an honest client. This parameter characterizes how system imperfections lead to rejection of the banknote in the verification phase wrongly. In \cite{bozzio2018experimental}, it has been shown that the lower bound for this parameter is given by
\begin{equation}
P_{\rm correct} \geq 1-e^{-cn \delta^2/2},
\label{P_correct}
\end{equation}
where $c$ is the probability of successful verification assuming an honest client, for mini-scheme quantum money protocol with just one qubit pair (for more  details  please  refer  to \cite{bozzio2018experimental}). The parameter $\delta$ is defined as
\begin{equation}
\delta=\frac{2c}{3}-\frac{7}{12}.
\end{equation}
It is worth noting that if $c>0.875$ the security of the protocol can be guaranteed.
\subsubsection*{Probability of forge}
Another figure of merit for this protocol is the probability of successful forging, e.g., duplication of the banknote by a dishonest client. An upper bound for this parameter is given by \cite{bozzio2018experimental}
\begin{equation}
P_{\rm forge}\leq e^{-n \delta^2/4}. 
\label{P_forge}
\end{equation}

\subsection*{Simulation results}
We use NetSquid to simulate the quantum money protocol explained above. In particular, our goal is to investigate the effect of quantum memory and measurement error on the performance of this protocol. Hence, we do not consider any source of loss. Simulation parameters are chosen based on currently achievable hardware parameters in nitrogen-vacancy (NV) center implementation. We assume T1T2 noise model with $T_1=10~{\rm hours}$ and $T_2=1{\rm s}$ \cite{coopmans2021netsquid}, {where $T_1$ and $T_2$ denote the decay time constant of the quantum memory and the decoherence time constant of the quantum memory, respectively}. Measurement error has been modelled with $p_1=0.05$ and $p_2=0.005$, where $p_1$ is the  probability that a measurement result $0$ is flipped to $1$, and $p_2$ is the probability that a measurement result $1$ is flipped to $0$ \cite{coopmans2021netsquid}. 

To evaluate the performance of the system, in the first step we obtain the parameter $c$ by simulation. The parameter $c$ is then used to calculate the security bounds for $P_{\rm correct}$ and $P_{\rm forge}$. To obtain $c$ by simulation, a block of $10000$ qubit pairs is sent and the simulation is repeated ten times to achieve better accuracy. The parameter $c$ is then given by $N_{\rm valid}/N_{\rm detected}$, where $N_{\rm valid}$ is the number of valid outcomes corresponding to the qubits in the chosen basis, and $N_{\rm detected}$ denotes the number of detected outcomes corresponding to the qubits in the chosen basis. 

\begin{figure}[h]
	\centering
	\includegraphics[scale=0.45]{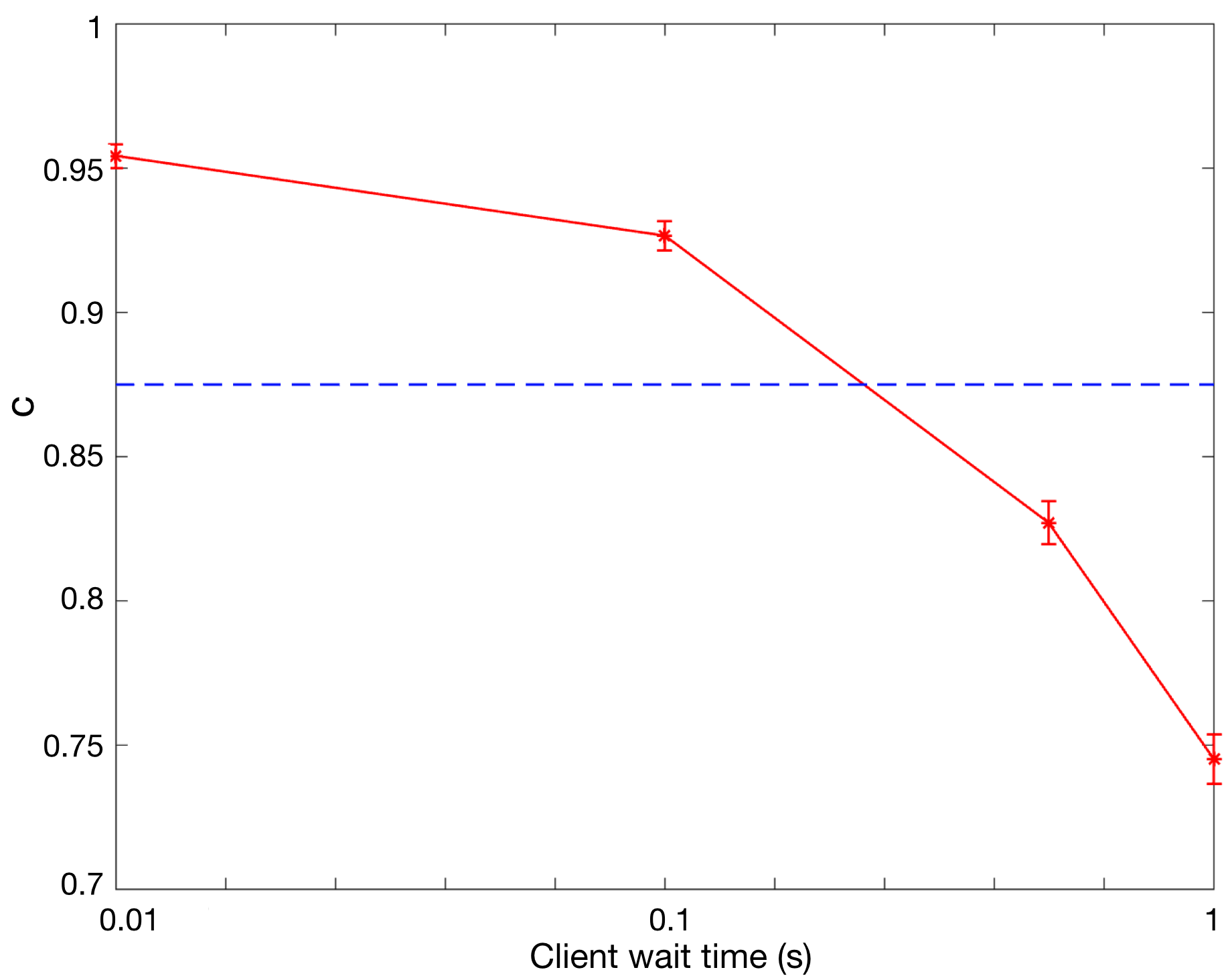}
	\caption{$c$ versus client wait time, $T$. The blue dashed line shows the security threshold $0.875$.
	} 
	\label{parameter_c}
\end{figure}

\begin{figure}[h]
	\centering
	\includegraphics[scale=0.65]{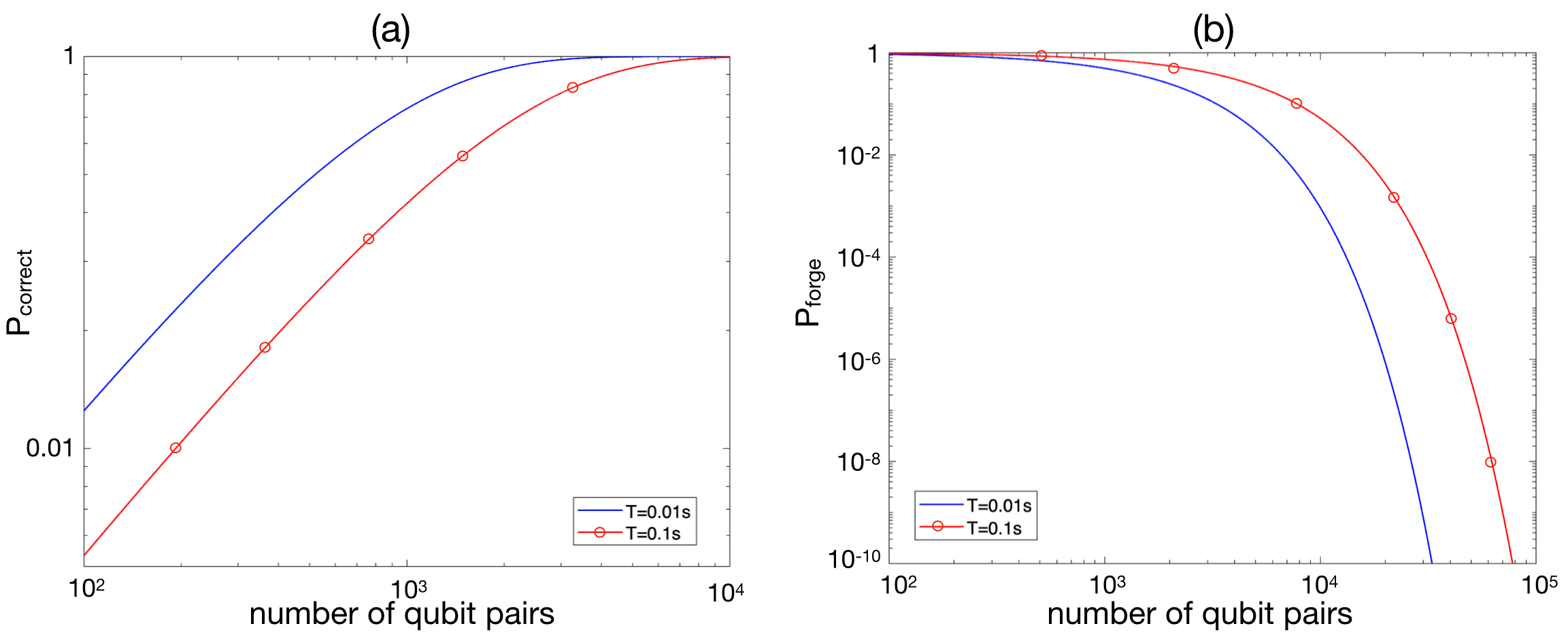}
	\caption{(a) $P_{\rm correct}$ versus number of qubit pairs for $T=0.01~{\rm s}$ and $T=0.1~{\rm s}$. (b) $P_{\rm forge}$ versus number of qubit pairs for $T=0.01~{\rm s}$ and $T=0.1~{\rm s}$.
	} 
	\label{Fig:P_correct_P_forge}
\end{figure}

Figure \ref{parameter_c} shows $c$ for different values of client wait time, $T$. The error bars corresponding to a confidence level of $95\%$ are also shown in the figure. The blue dashed line shows the threshold $0.875$. The parameter $T$ characterizes any delay in the system before measuring the qubits; hence it is equal to the storage time of qubits in the quantum memory. {Note that for a quantum memory with parameters $T_1$ and $T_2$, a longer storage time results in a higher decoherence noise and a lower $c$.} It can be seen that for $T$ larger than about $0.3~{\rm s}$ the protocol is not guaranteed to be secure anymore. This value is about one third of $T_2=1s$. This shows that minimizing the storage time of qubits in the quantum memory is crucial in the practical implementation of this protocol. 
In order to evaluate the required number of qubit pairs to achieve a specific security level, the parameters $P_{\rm correct}$ and $P_{\rm forge}$, for $T=0.01~{\rm s}$ and $T=0.1~{\rm s}$ are shown in figure \ref{Fig:P_correct_P_forge}. According to Fig.~\ref{parameter_c}, the security of the protocol is guaranteed in these two values of $T$. We consider the threshold $10^{-7}$ for $P_{\rm forge}$. To achieve this threshold, the number of qubit pairs is required to be larger than about $n=2.3\times 10^{4}$ for $T=0.01~{\rm s}$, whereas for $T=0.1~{\rm s}$, the minimum required $n$ increases to $5.4\times 10^{4}$. This implies that the storage time of qubits has a huge effect on the minimum qubit pairs required to achieve a specific security level. For more details on NetSquid simulation for this protocol please refer to Appendix~D (Supplementary Information). 


\section*{Benchmarking of W-state based Anonymous Transmission}
\label{Sec_AT}
Anonymous transmission addresses the issue of concealing the identity of two communicating nodes in a quantum network with $N$ nodes. More specifically, the identity of the sender $S$ is required to be unknown to all other nodes in the network, whereas the identity of the receiver $R$ is hidden to all other parties except the sender. In this section, we consider W-state based anonymous transmission protocol \cite{lipinska2018anonymous}. This protocol is mainly based on the establishment of anonymous entanglement between $S$ and $R$. The entangled state between $S$ and $R$ is then used to teleport the desired state.

\begin{figure*}[t]
	\centering
	\includegraphics[scale=0.84]{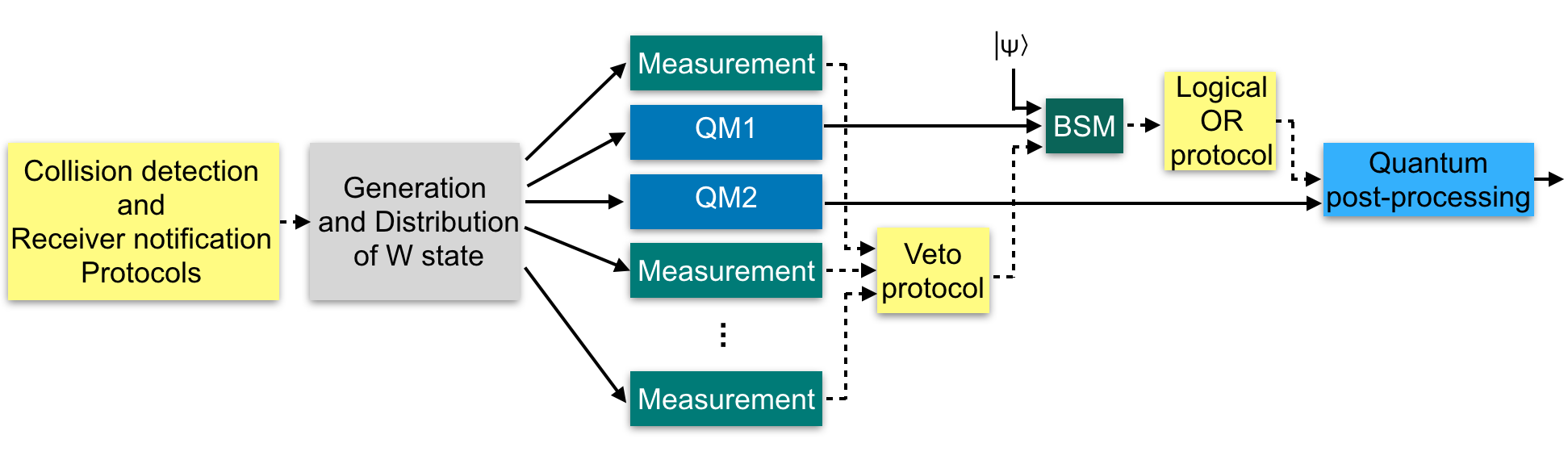}
	\caption{General description of W-state based anonymous transmission protocol. Solid arrows represent quantum communication, while dotted arrows represent classical communication. QM: quantum memory, BSM: Bell state measurement. 
	} 
	\label{model_AT}
\end{figure*}

Figure \ref{model_AT} shows the general description of this protocol. At the first step, collision detection and receiver notification protocols \cite{broadbent2007information} are used to determine a single sender $S$ and notify the receiver $R$, respectively. Then, W state is generated and distributed among the users. In the next step, all users except for $S$ and $R$ perform measurement in the standard basis, while $S$ and $R$ keep their particles in the quantum memory. Measurement outcomes are, then, used in veto protocol \cite{broadbent2007information}, which determines whether all $N-2$ outcomes are zero or not. In the latter case, the protocol aborts, while in the former case, it is assumed that anonymous entanglement is established between $S$ and $R$. In that case, the protocol proceeds with the teleportation of the state $\left|\psi\right\rangle$. To this aim, $S$ performs Bell state measurement and sends the two classical outcomes anonymously to $R$ using logical OR protocol. The receiver $R$, then, uses this information to perform suitable quantum post-processing on his qubit. 

To evaluate the performance of the protocol in the presence of system imperfections, it is necessary to consider major sources of loss and noise in the system, as summarized in Table~\ref{AT_noise}. One major source of noise is the delay caused by classical sub-algorithms. During the run time of veto protocol, the particles of $S$ and $R$ are stored in the quantum memory, which introduces decoherence to the particles. Quantum memory noise also affects $R$'s particle during the run time of logical OR protocol. Imperfect operation of quantum gates in the quantum post-processing step also introduces some noise in the system. Another source of noise is nonideal generation and distribution of W state, which is investigated in \cite{lipinska2018anonymous}. Aside from these nonidealities, losses in the system such as transmission loss (e.g. optical fibre loss) and the loss introduced by quantum memory may also adversely affect the performance of the protocol. 

In the following, we consider several figures of merit for this protocol and investigate them in more details.

\begin{table*}[t]
	\caption{Major sources of noise and loss in anonymous transmission protocol.}
	\centering 
	\label{noise_loss_AT}
	\begin{tabular}{|c |c|} 
		\hline
		\hline
		Protocol Step & Major sources of noise and loss \\
		\hline 
		\hline
		\multirow{2}{*}{generation and distribution of W state} & decoherence \\
		\cline{2-2} 
		&  transmission loss\\
		\hline
		measurement performed by $N-2$ users &   loss introduced by measurement \\
		\hline
		\multirow{2}{*}{veto protocol}& noise introduced by quantum memory\\ 
		\cline{2-2}
		& loss introduced by quantum memory\\
		\hline
		Bell state measurement & loss introduced by measurement \\
		\hline
		\multirow{2}{*}{anonymous transmission of two classical bits} & noise introduced by quantum memory \\ 
		\cline{2-2}
		& loss introduced by quantum memory \\
		\hline
		quantum post-processing at receiver side & noise introduced by quantum gates\\
		\hline
		\end{tabular}
	\label{AT_noise}
\end{table*} 

\subsection*{Figures of merit}

\subsubsection*{probability of protocol failure}
W-state based anonymous transmission protocol is probabilistic \cite{lipinska2018anonymous}, i.e., anonymous entanglement is established between $S$ and $R$ with some probability. Aside from that, losses in the system may cause the protocol to fail. In \cite{lipinska2018anonymous}, it has been shown that this protocol tolerates one nonresponsive node among $N-2$ nodes (all nodes except $S$ and $R$). Nevertheless, if $S$ or $R$ lose their particle, the protocol will fail. We can write the probability of protocol failure as follows:
\begin{eqnarray}
    P_{\rm fail}=(1-Pr(D)+Pr(D)Pr(A|B))Pr(B)\nonumber\\
    +(1-Pr(D)+Pr(D)Pr(A|C))Pr(C)\nonumber\\
    +(1-Pr(B)-Pr(C)),\quad \quad\quad\quad\quad\quad\;
    \label{p_failure}
\end{eqnarray}
where $A$ is the event that at least one of measurement outcomes (inputs of veto protocol) is not zero, $B$ is the event that all $N-2$ nodes that perform measurement are responsive, $C$ is the event that just one out of $N-2$ nodes that perform measurement is nonresponsive, and $D$ is the event that both $S$ and $R$ does not lose their particle. The probability $Pr(B)$ can be obtained by
\begin{equation}
Pr(B)=\prod_{\begin{array}{c}i\in \{1,...,N\}\\
i\neq i_S,i_R\end{array}} {\eta_i},
\end{equation}
where $\eta_i$ is the transmittance corresponding to the $i$th node. Similarly, the probability $Pr(C)$ can be expressed as
\begin{equation}
    Pr(C)=\sum_{i=1, i\neq i_S,i_R}^{N}{\{(1-\eta_i)\prod_{\begin{array}{c}j\in \{1,...,N\}\\
j\neq i_S,i_R,i\end{array}} {\eta_j}\}}.
\end{equation}
The parameter $Pr(D)$ is given by
\begin{equation}
    Pr(D)=\eta_{\rm BSM}\eta_{i_S}\eta_{i_R},
\end{equation}
where $\eta_{\rm BSM}$ denotes the loss introduced by Bell state measurement. The probabilities $Pr(A|B)$ and $Pr(A|C)$ are calculated in Appendix~A (Supplementary Information).
\subsubsection*{probability of correctness}
Probability of correctness is defined as the probability of successful teleportation of the state $\left | \psi\right\rangle$, under the assumption that all participants are honest and the protocol does not fail. We can write this parameter as follows:
\begin{equation}
P_{\rm correct}=(1-\epsilon_{\rm corr}).
\end{equation}
In the above equation, $\epsilon_{\rm corr}$ represents the probability of failure in classical subroutines \cite{lipinska2018anonymous}. 
\subsubsection*{Fidelity of anonymous entanglement between $S$ and $R$}
The fidelity of the entangled state established between $S$ and $R$ is another important figure of merit for this protocol. This parameter is defined as 
\begin{equation}
  F(\gamma)=Tr[\gamma . \left | \Psi^{+} \right \rangle \left \langle \Psi ^{+} \right |],  
\end{equation}
where $\gamma$ is the anonymous entangled state between $S$ and $R$ and $\left | \Psi^{+} \right \rangle= \frac{1}{\sqrt{2}}(\left | 01 \right \rangle+\left | 10 \right \rangle)$.
\subsubsection*{Average fidelity of the teleported state}
The goal of anonymous transmission protocol is to transmit a quantum state anonymously. Therefore, one of the most important figure of merits is the quality of the teleported state, which can be characterized by average fidelity. We write the quantum state to be teleported as a Bloch vector in Bloch sphere as follows:
\begin{equation}
|\psi\rangle=\cos \left(\frac{\theta}{2}\right)e^{i\phi/2}|0\rangle+\sin \left(\frac{\theta}{2}\right)e^{-i\phi/2}|1\rangle,
\end{equation}
where $\theta$ and $\phi$ denote the polar and azimuthal angles, respectively. The average fidelity of the teleported state is given by \cite{oh2002fidelity}
\begin{equation}
F_{\rm ave}=\frac{1}{4 \pi}\int_{0}^{\pi}d\theta\int_{0}^{2\pi}{F(\theta,\phi)\sin(\theta) d\phi},
\label{fidelity_ave}
\end{equation}
where
\begin{equation}
    F(\theta,\phi)={\rm Tr}[|\psi\rangle\langle \psi|\rho_{out}]
\end{equation}
In the above equation, $\rho_{out}$ denotes the density matrix for the teleported state corresponding to $|\psi\rangle$.

\subsection*{Simulation results}
In this section, we present some simulation results for anonymous transmission protocol. We use NetSquid to simulate this protocol for four users. The average fidelity of the teleported state (equation (\ref{fidelity_ave})) is approximated using Reimann sum as follows:
\begin{equation}
  F_{\rm ave}  \simeq  \frac{\pi}{2\times6400}\sum_{k=0}^{79}{\sum_{m=0}^{79}{g(\theta_k,\phi_m)}},
\end{equation}
where $\theta_k=\frac{\pi}{160}+k\frac{\pi}{80}$, and $\phi_m=\frac{\pi}{80}+m\frac{2\pi}{80}$, for $k=0,1,2,...,79$ and $m=0,1,2,...,79$. 

First of all, the effect of the noise introduced by quantum memories on the quality of the teleported state is evaluated. We assume that the particles of $S$ and $R$ are stored for the time duration of $t_1$ before teleportation. $R$'s particle is assumed to be kept in the quantum memory for an additional time interval of $t_2$, i.e., total storage time for $R$'s particle is $t_1+t_2$. We consider dephasing noise model for quantum memories. The dephasing probabilities for $S$'s and $R$'s quantum memories are denoted by $q_1$ and $q_2$, respectively. The parameters $q_1$ and $q_2$ correspond to $t_1$ and $t_1+t_2$, respectively. 

\begin{figure}[t]
	\centering
	\includegraphics[scale=0.53]{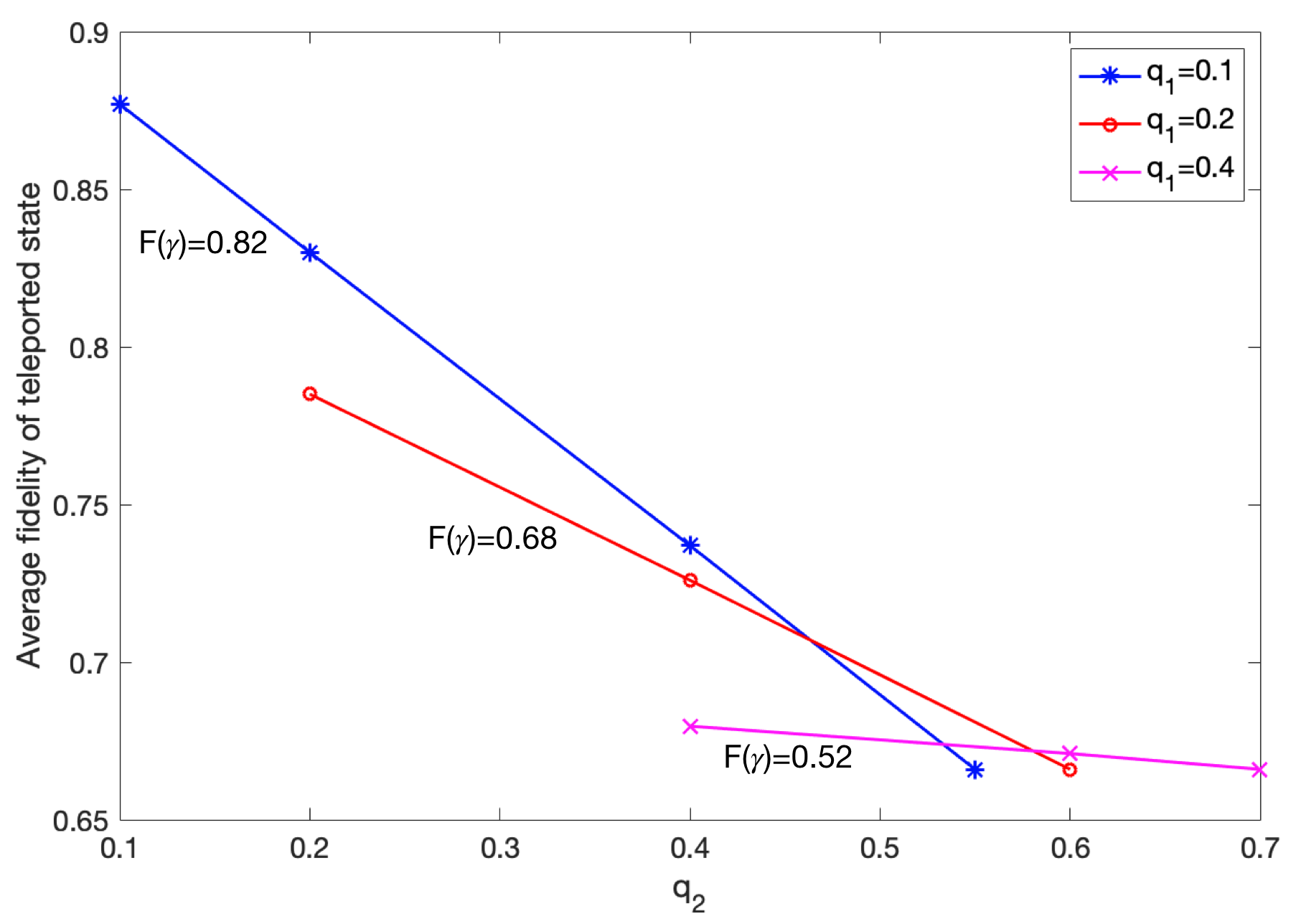}
	\caption{Average fidelity of the teleported state for different values of $q_1$ and $q_2$.
	} 
	\label{Ave_fidelity_memory}
\end{figure}

Figure~\ref{Ave_fidelity_memory} shows $F_{\rm ave}$ for different values of $q_1$ and $q_2$. The fidelity of the anonymous entangled state, $F(\gamma)$, corresponding to each value of $q_1$ is also obtained by \cite{lipinska2018anonymous}
\begin{equation}
 F(\gamma)=1-2q_1(1-q_1),   
\end{equation}
and shown in Fig.~\ref{Ave_fidelity_memory}. It can be seen that for $q_1=0.2$, the average fidelity of teleported state is already less than $0.8$. With the assumption of $q_1=(1-e^{-t_1/T_2})/2$, where $T_2$ is the decoherence time constant of the quantum memory, this value corresponds to $t_1/T_2=0.51$. This implies that it is crucial to minimize the delay caused by veto protocol as much as possible. 

It is worth noting that if $q_2=q_1$, there is no noise after Bell state measurement. In this case, we can analytically calculate $F_{\rm ave}$ from $F(\gamma)$ using the formula $F_{\rm ave}=(2F(\gamma)+1)/3$ \cite{horodecki1999general}. It can simply be concluded that the analytical results obtained by this formula validates the simulations results for the cases $q_2=q_1$ in Fig.~\ref{Ave_fidelity_memory}.

\begin{figure}[t]
	\centering
	\includegraphics[scale=0.78]{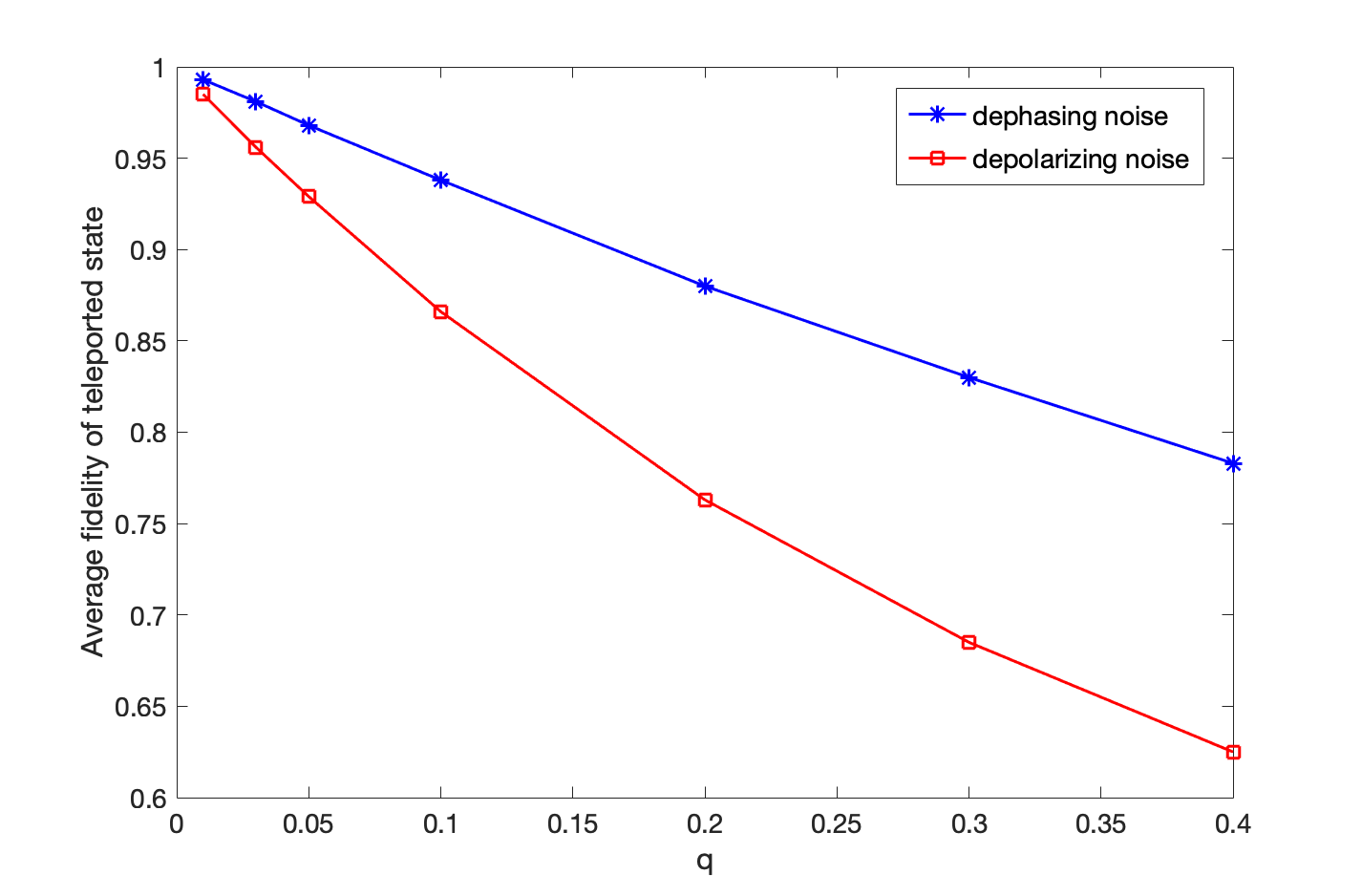}
	\caption{Average fidelity of the teleported state versus $q$.
	} 
	\label{ave_fidelity_gates}
\end{figure}

Next, we evaluate the performance of the protocol in the presence of noise at the $X$ and $Z$ gates in the final step of the teleportation. We consider two noise models, dephasing and depolarizing, with dephasing/depolarizing probability denoted by $q$. Figure \ref{ave_fidelity_gates} shows $F_{\rm ave}$ versus $q$. It can be seen that the average fidelity of the teleported state in case of dephasing noise model is significantly higher than that of depolarizing noise model, especially for large values of $q$. As an example, in case of dephasing noise, for $q$ smaller than about $0.15$, $F_{\rm ave}$ is above $0.9$, whereas in case of depolarizing noise we achieve this performance for $q$ smaller than about $0.075$. It is worth noting that the results presented in this section are independent of the number of users, $N$, since the sources of noise after measurement by all users except $S$ and $R$ are considered. For more details on NetSquid simulation for this protocol please refer to Appendix~D (Supplementary Information). 

Lastly, we present a numerical example to examine the effect of sources of loss in the system on the probability of protocol failure. We assume $\eta_{\rm d}=\eta_{\rm BSM}=0.8$, where $\eta_{\rm d}$ denotes the measurement loss for each of $N-2$ users. The transmittance corresponding to $i$th node (except $S$ and $R$ is, then, assumed to be $\eta_i=\eta_{\rm d}\eta_{\rm tr}$), where $\eta_{\rm tr}$ represents transmission loss. The loss introduced by quantum memories is modelled by $\eta_{\rm qm}=\eta_{0}e^{-t_{\rm s}/T_1}$. Here, $T_1$ denotes decay time constant of quantum memory, $t_{\rm s}$ denotes the storage time, and $\eta_{0}$ is a constant less than one. The parameter $\eta_{0}$ is assumed to be $0.8$. The ratio $t_{\rm s}/T_1$ is assumed to be $0.002$ for sender,  and $0.004$ for receiver.

\begin{figure}[h]
	\centering
	\includegraphics[scale=0.8]{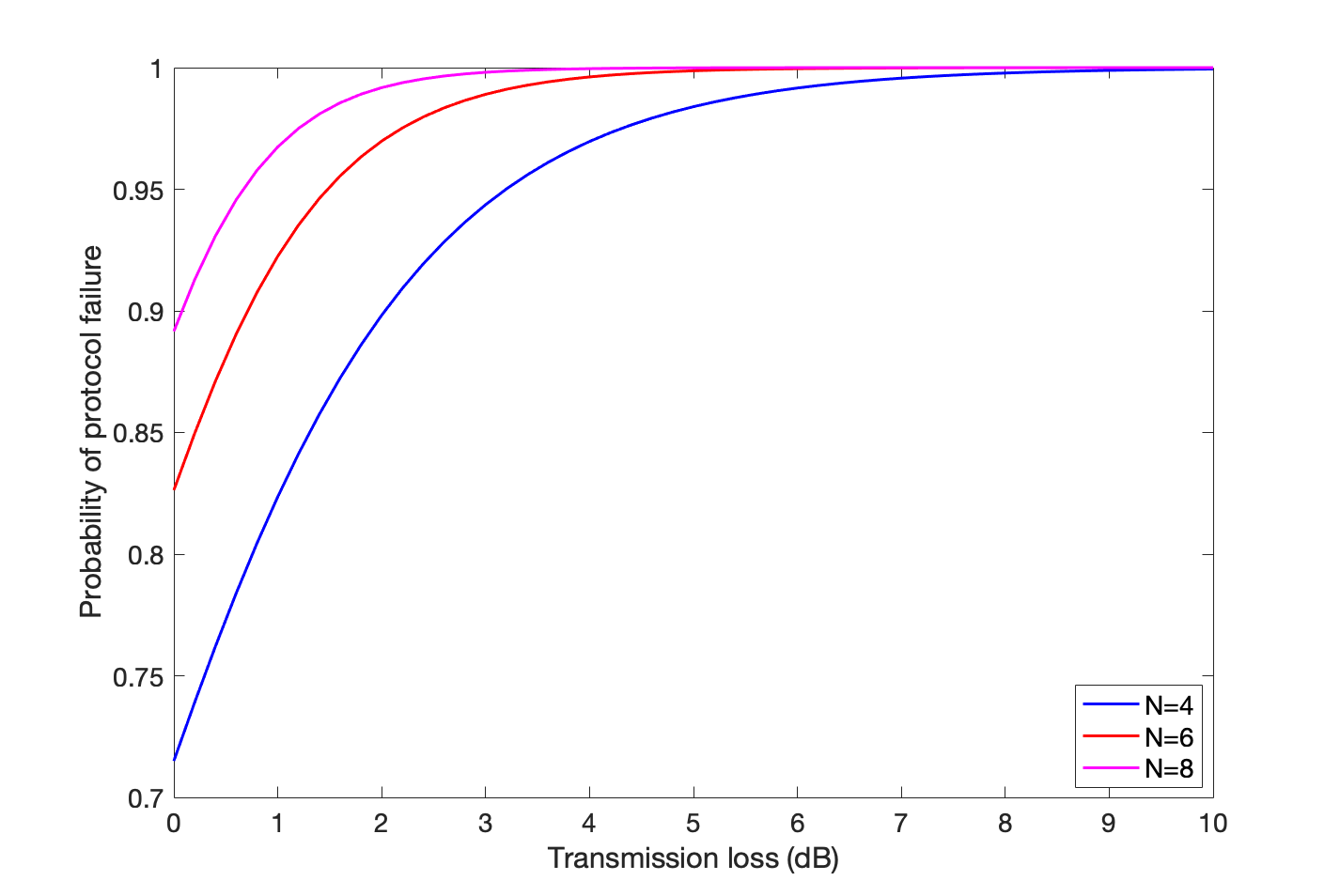}
	\caption{Probability of protocol failure versus transmission loss for different number of users.
	} 
	\label{AT_failure}
\end{figure}

Figure \ref{AT_failure} shows the probability of protocol failure versus transmission loss for different values of $N$. It can be seen that the probability of protocol failure increases with the increase of number of users. For $N=4$, this probability reaches one for transmission of loss about $10~{\rm dB}$. This value reduces to about $8~{\rm dB}$ and $5.5~{\rm dB}$ for 6 and 8 users, respectively. 

In Appendix~A (Supplementary Information), two cases of ideal W state  and noisy W state with dephasing noise model are considered. It has been shown that dephasing noise does not change the probabilities $Pr(A|B)$ and $Pr(A|C)$ in (\ref{p_failure}) compared to that of noiseless case. Hence, the presented results in Fig.~\ref{AT_failure} are applicable to noisy W state with dephasing noise model as well.
\section*{Benchmarking of verifiable blind quantum computation}
\label{Sec_VBQC}
VBQC enables delegating quantum computation to a quantum server while preserving privacy \cite{kashefi2020vbqc}. In this paper, we choose the measurement-based VBQC protocol proposed in \cite{kashefi2020vbqc}. The steps of the protocol are outlined in Appendix~B (Supplementary Information). We assume that three qubits are used at the server side. This protocol consists of $d$ computation runs and $t$ test runs. If the number of failed test runs is larger than a threshold denoted by $w$, the protocol aborts. 

The main sources of noise in this protocol are imperfect operation of quantum gates and measurement errors. Aside from that, classical communication between client and server may introduce some delay, which substantially causes decoherence. Lastly, transmission loss may be troublesome, depending on the distance between client and server. In the following, two figures of merit for this protocol are presented.

\subsection*{Figures of merit}
The probability of aborting assuming an honest server is one of the main figure of merits for this protocol. Suitable choice of protocol parameters such as $w$ and $t$ plays an important role in avoiding unnecessary aborting of the protocol while maintaining the security.

Another figure of merit for this protocol is probability of correctness, which is defined as the probability of correct output assuming the protocol does not abort. 

\subsection*{Simulation results}
In this subsection, we examine the performance of VBQC protocol in the presence of system imperfections. We use NetSquid to simulate the VBQC protocol outlined in Appendix~B (Supplementary Information). We assume that the server has three qubits. Nominal values used for time duration of gates are listed in Table \ref{Tab_VBQC}. These parameters are chosen based on the NV platform implementation \cite{coopmans2021netsquid}. We assume depolarizing noise model for the quantum gates. In our simulation, we assume that the depolarizing probability in quantum gates are identical. This provides us a benchmark for the performance of the protocol. As for measurement error, we use the bit flip model used in \cite{coopmans2021netsquid}. We assume that the measurement outcome 0 is flipped to 1 with probability $0.05$, whereas the measurement outcome 1 is flipped to 0 with probability $0.005$.    

\begin{figure}[t]
	\centering
	\includegraphics[scale=0.75]{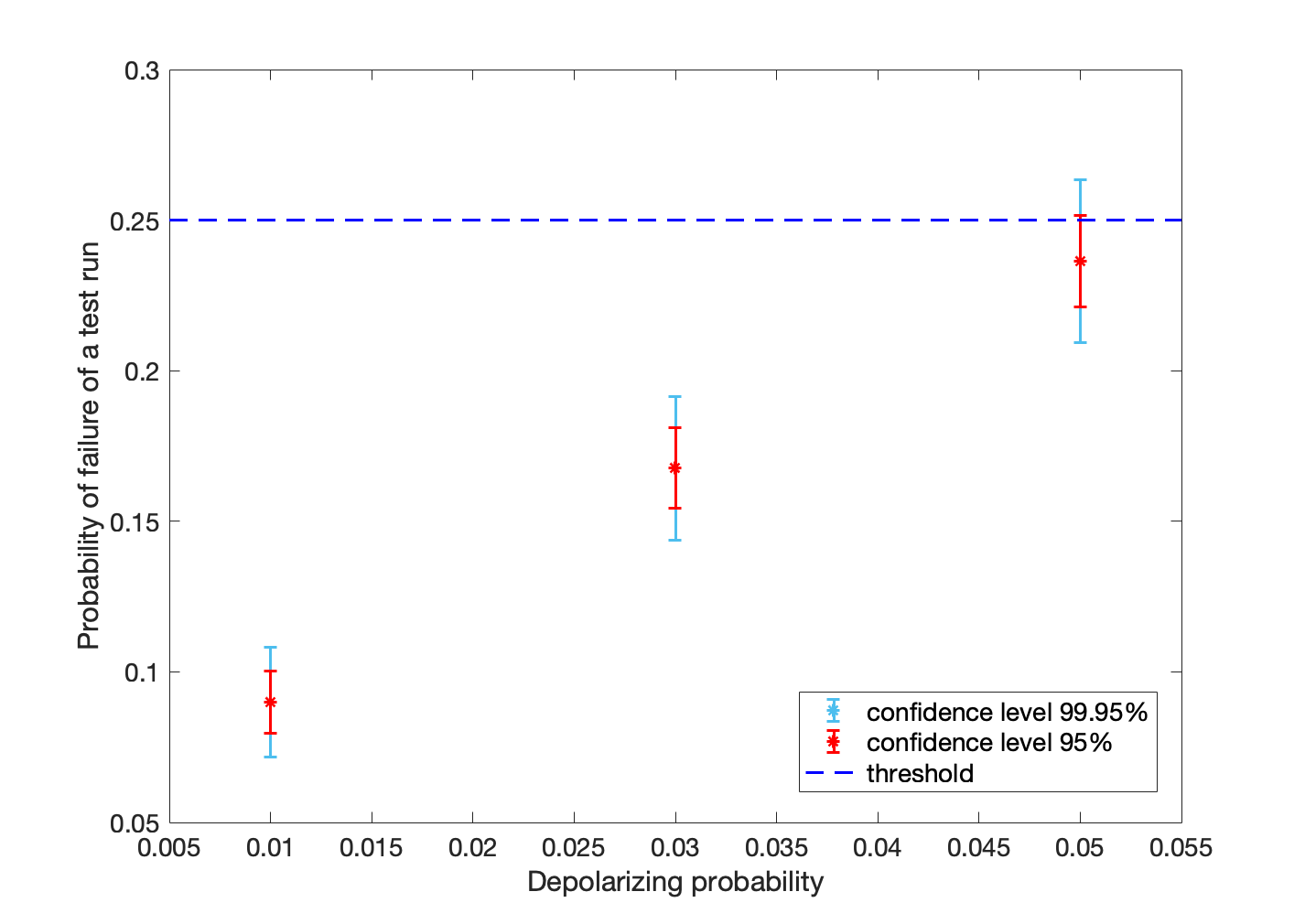}
	\caption{Probability of failure of a test run for different values of depolarizing probability in quantum gates. The blue dashed line represents the threshold $0.25$ for $w/t$. 
	} 
	\label{vbqc}
\end{figure}

First of all, we evaluate the performance of a test run. We consider different values for depolarizing probability in quantum gates. In each case, the test run is performed for $3000$ times and the probability of failure of a test run is calculated, as shown in Fig.~\ref{vbqc}. The upper and lower bounds for this probability, denoted by $P_{\rm max}$ and $P_{\rm min}$, respectively, depend on the desired confidence level. Here, the results for two confidence levels $95 \%$ and $99.95 \%$ are shown in Fig.~\ref{vbqc}. 

We use the obtained values for $P_{\rm max}$ and $P_{\rm min}$ to determine the suitable range of values for $w/t$. In \cite{kashefi2020vbqc}, it has been shown that if $w/t > P_{\rm max}$, the protocol is $\epsilon_{\rm c}$-locally-correct with exponentially low $\epsilon_{\rm c}$, i.e., with honest parties the output will be the expected one. On the other hand, according to \cite{kashefi2020vbqc}, in order for the protocol to be secure, $w/t$ should be less than $1/2k$, where $k$ is the number of colouring in the protocol (for more information please refer to \cite{kashefi2020vbqc}). In our simulated protocol with a three-qubit server, $k=2$. Hence, we have $P_{\rm max}<w/t<0.25$. The threshold $1/2k=0.25$ is shown in Fig.~\ref{vbqc} by the blue dashed line. It can be seen from fig.~\ref{vbqc} that with our chosen parameters, for a depolarizing probability of $0.05$, there is no acceptable value for $w/t$. If the depolarization probability is reduced to $0.03$, it is feasible to choose parameter values satisfying $P_{\rm max}<w/t<0.25$. Nevertheless, there are limited acceptable values for parameters $w$ and $t$. To extend the range of acceptable parameter values and have more flexibility in choice of protocol parameters, it is necessary to enhance the depolarization probability of quantum gates beyond $0.03$. 
\begin{table}[b]
	\caption{Time duration of gates used in VBQC protocol}
	\centering 
	\begin{tabular}{|c |c |} 
		\hline
		&duration \\
		\hline		
		single-qubit gate& $5~{\rm ns}$\\
		CNOT gate& $20~{\rm \mu s}$\\
		Control Z gate & $20~{\rm \mu s}$\\
		measurement & $3.7~{\rm \mu s}$\\
		\hline
	\end{tabular}
	\label{Tab_VBQC}
\end{table} 

In order to obtain the optimum values for the protocol parameters $w$, $t$, and $d$, it is required to solve the optimization problem mentioned in \cite{kashefi2020vbqc}, which is beyond the scope of this paper. Nevertheless, we provide an example by considering a depolarizing probability of $0.03$ and $w=1$. With these values, the only acceptable value for the number of test runs is $t=5$. We run the protocol for $3000$ times with $5$ test runs and $6$ computation runs. The resulting probability of correctness is $0.929\pm 0.0092$, considering a confidence level of $95\%$. For more details on NetSquid simulation for this protocol please refer to Appendix~D (Supplementary Information). 

\section*{Benchmarking of quantum digital signature}
\label{SEc_QDS}
\label{Sec_QDS}
QDS schemes can guarantee the unforgeability, nonrepudiation, and transferability of a signed message with information-theoretical security. Here, we consider the three-party QDS protocol proposed in \cite{amiri2016secure}. The protocol is outlined Appendix~C (Supplementary Information). 

In the practical implementation of this protocol, various imperfections such as transmission loss and measurement error may adversely affect the performance of the protocol. In the following, the figures of merit for this protocol are discussed.

\subsection*{Figures of merit}
We can define three main figures of merit for the security of this protocol. First of all, the probability of aborting under the assumption that all three parties are honest characterizes the robustness of the protocol. Another figure of merit is the probability of successful forging the signature by Bob. The third security criteria is the probability of repudiation, i.e., the probability that the signature sent by Alice is accepted by Bob, but when Bob forwards it to Charlie it is rejected. We denote the security bounds on these three probabilities by $P_{\rm abort}$, $P_{\rm for}$, and $P_{\rm rep}$, respectively. The security level of the protocol is then defined as $\beta=\max \{P_{\rm abort}, P_{\rm for}, P_{\rm rep}\}$. 

\subsection*{Simulation results}
In this subsection, the simulation results for the quantum digital signature protocol described in Appendix~C (Supplementary Information) are presented. The number of transmitted qubits is assumed to be $5\times 10^{4}$. We assume Alice and Bob are connected via an optical fibre with length $L_{\rm fib}$. Similarly, the distance between Alice and Charlie is $L_{\rm fib}$. The total loss of the system (excluding the loss of the optical fibre) is assumed to be $0.5$. The attenuation coefficient of the optical fibre is $0.2~{\rm dB/km}$. We set the protocol parameters $\epsilon$, $\epsilon_{PE}$, $a$ and r as $\epsilon=10^{-10}$, $\epsilon_{PE}=10^{-5}$, $a=10^{-5}$, and $r=0.1$. Here, $\epsilon_{PE}$ is the failure probability in calculating the upper bounds for error rates, $r$ is the fraction of sifted key used in error estimation, and $a$ and $\epsilon$ are small constants. For more information on these parameters please refer to Appendix~C (Supplementary Information). 

We consider different values for the length of optical fibre and evaluate the performance of the system in the presence of channel loss. Furthermore, two cases for the error in the measurement setup, denoted by $e_d$, are considered. The first case is the ideal case where there is no error, whereas in the second one $e_{\rm d}=0.015$.

According to equation (8) in Appendix C, we have $P_{\rm abort}=2\epsilon_{PE}=2\times 10^{-5}$. As for $P_{\rm for}$, our results show that $P_{\rm for}=10^{-4}$ for all cases. This is mainly because for our chosen parameter values, $l$ is sufficiently large such that $\epsilon_F \simeq \epsilon/a=10^{-5}$. 

Table \ref{QDS_results} shows the parameter $P_{\rm rep}$ for different values of $L_{\rm fib}$ and $e_d$. It can be seen that as fibre length increases, the security bound for probability of repudiation increases significantly. This mainly happens due to the reduction of the number of received signals, which results in a shorter signature. Hence, $P_{\rm rep}$ increases according to equation (9) in Appendix C (Supplementary Information). Additionally, a shorter raw key leads to more statistical fluctuations in the estimation of error parameters (see Appendix~C (Supplementary Information)), which results in a larger $P_{\rm rep}$. For more details on NetSquid simulation for this protocol please refer to Appendix~D (Supplementary Information).  
\begin{table}[h]
	\caption{$P_{\rm rep}$ for different values of $L_{\rm fib}$ and $e_d$}
	\centering
	\label{QDS_results}
	\begin{tabular}{|c |c|} 
		\hline
Parameter values & $P_{\rm rep}$\\
		\hline
$L_{\rm fib}=5~{\rm km}$,$e_d=0$& $6.7 \times 10^{-5}$\\
\hline
$L_{\rm fib}=10~{\rm km}$,$e_d=0$& $0.008$\\
\hline
$L_{\rm fib}=20$,$e_d=0$& $0.7$\\
\hline
$L_{\rm fib}=5$,$e_d=0.015$& $0.1927$\\
\hline
$L_{\rm fib}=10$,$e_d=0.015$& $0.698$\\
\hline
$L_{\rm fib}=20$,$e_d=0.015$& $1$\\
\hline
	\end{tabular}
\end{table}

\section*{Backward benchmarking}
In the previous subsections, we presented a benchmarking method for quantum protocols based on determining fixed values for system parameters and evaluating the performance of the protocol considering various figures of merit. Another method to benchmark quantum protocols is to consider target values for figures of merit and optimize system parameters with the aim of minimizing hardware requirements while satisfying the target figures of merit. We refer to this method as "backward benchmarking``. In \cite{da2021optimizing}, a method based on this type of benchmarking has been proposed to optimize entanglement generation and distribution in quantum networks using genetic algorithms. By appropriately redesigning the cost function used, this method can be adapted to evaluate the performance of various other quantum protocols and optimize hardware performance.

In this paper, as an example, we use the method proposed in \cite{da2021optimizing} to benchmark the quantum money protocol. Specifically, our goal is to determine minimum requirements to guarantee the security of this protocol, i.e., $c>0.875$, considering a fixed value for the storage time of qubits in the quantum memory. We are interested in answering the question of what are minimum viable improvements required for the quantum memory parameters $T_1$ and $T_2$, compared to their currently achievable values experimentally (referred to as "baseline values`` in \cite{da2021optimizing}), to achieve $c>0.875$ for a desired fixed storage time. 

To obtain the minimum requirements for a specific storage time, we consider an optimization problem with the following cost function
\begin{equation}
T_c=w_1 \Theta (c_{min}-c)+w_2 C(T^{\prime}_1,T^{\prime}_2),
\label{cost}
\end{equation}
where the parameters $w_1$ and $w_2$ are the weights of the objectives in the cost function, $\Theta(\cdot)$ is the step function, and 
\begin{equation}
C(T^{\prime}_1,T^{\prime}_2)=\frac{1}{\log_{T^{\prime}_{1b}}T^{\prime}_1}+\frac{1}{\log_{T^{\prime}_{2b}}T^{\prime}_2}.
\label{cost_func}
\end{equation}
In the above equation, the parameters $T^{\prime}_i$ and $T^{\prime}_{ib}$, for $i=1,2$, are in the range $[0,1]$. Hence, $T_i$ and $T_{ib}$ are converted to a value in this range using the following equations:
\begin{eqnarray}
T^{\prime}_i=\frac{T_i}{1+T_i},\nonumber\\
T^{\prime}_{ib}=\frac{T_{ib}}{1+T_{ib}}.
\end{eqnarray}
By ensuring that both the base and the argument of the logarithm in (\ref{cost_func}) are in the range $[0,1]$, we guarantee that the cost function reflects the progressive hardness of improving hardware parameters. By this we mean that to improve $T_i$ slightly over $T_{ib}$, only a small cost is assigned. However, as $T_i$ gets closer to its perfect value, the cost grows to infinity. This is meant to reflect the expectation that in an experimental setting, a hardware parameter becomes more difficult to improve as it approaches its perfect value. For further discussion on this point, see \cite{da2021optimizing}. We employ a genetic algorithm-based optimization methodology to minimize the cost function in (\ref{cost}). For more information on the optimization procedure, please refer to \cite{da2021optimizing}. 

\begin{table*}[t]
	\caption{Optimum solutions for values of $T_1$ and $T_2$.}
	\centering 
	\label{optimization_results}
	\begin{tabular}{|c |c|c|} 
		\hline
		\hline
		Storage time (s) & $T_1$ (h) & $T_2$ (s) \\
		\hline 
		\hline
		1& 10.037 & 3.25\\
		\hline
		2 & 10.05 & 6.21\\
		\hline
		5 & 10.099 & 16.007\\
		\hline
		\end{tabular}
\end{table*} 

As in Sec.~2, NetSquid is used to simulate the quantum money protocol. To obtain the parameter $c$, we use a block for 1000 qubit pairs and repeat the simulation for 5 times. For the baseline values, we choose $T_{1b}=10~\rm{h}$ and $T_{2b}=1s$. Other system parameters are the same as the ones chosen in Sec.~2. Table \ref{optimization_results} shows the optimum solutions for values of $T_1$ and $T_2$. Comparing these values with the baseline values, it can be seen that $T_2$ requires much more improvement than $T_1$. This confirms that the main parameter limiting the performance of quantum money protocol is $T_2$. Another observation is that the optimal solution for $T_2$ is around three times the storage time of qubits in the quantum memory, which is in line with the results obtained in Sec.~2. 
\section*{Conclusion and future outlook}
\label{Sec_con}
Quantum protocols enable distinctive functionalities such as secure communications and blind computation. To determine the requirements of quantum protocols and benchmark them against classical and post-quantum protocols, it is necessary to evaluate their performance considering different sources of system imperfection. Here, we considered several quantum protocols, namely quantum money, W-state based anonymous transmission, verifiable blind quantum computation, and quantum digital signature. We performed in-depth performance analysis for each protocol, mainly by use of NetSquid simulation platform. 

First, we examined the effect of decoherence noise introduced by quantum memory in quantum money protocol. Our simulation results showed that the coherence time of quantum memory is the main parameter limiting the practical implementation of this protocol. To guarantee the security of this protocol, the coherence time of quantum memory is required to be at least three times the storage time of qubits. To enable longer than one second storage time while guaranteeing the security, it is necessary to improve hardware parameters, especially the coherence time of quantum memory.  

Next, we considered W-state based anonymous transmission protocol and examined the degrading effect of different nonidealities such as decoherence noise of quantum memory, loss, and gate imperfections in the teleportation step. It can be inferred from the simulation results that the decoherence time constant of the quantum memory and the storage time of quantum particles play an important role in the fidelity of the teleported state. {For instance, to achieve an average fidelity above 0.8 in this protocol, the storage time of the sender's and receiver's particles in the quantum memory must be less than half of the decoherence time constant of the quantum memory. This implies that aside from improving hardware parameters such as the decoherence time constant of quantum memory, minimizing the delay caused by the classical sub-protocols used in this protocol is of paramount importance. This can be achieved by using high-speed processors for the classical sub-protocols, and reducing the number of times the parity sub-protocol is repeated in the veto protocol, denoted by $s$. According to \cite{broadbent2007information}, the correctness of the veto protocol decreases with reducing $s$. This imposes a trade-off between the correctness of the anonymous transmission protocol and the fidelity of the teleported state. The simulation results presented in this work are a great tool for efficient choice of protocol and hardware parameters, especially the parameters of classical sub-protocols such as $s$.}

{We also evaluated the degrading effect of the transmission loss on the probability of protocol failure. Although W-state based anonymous transmission protocol is more robust to loss of particles compared to its GHZ-based counterpart, our numerical results show that the probability of failure significantly increases with loss. Further, it was shown that the probability of protocol failure increases significantly by increasing the number of protocol participants, $N$. For instance, with our chosen parameters, for a transmission loss of 1 dB the probability of protocol failure for $N=4$ and $N=8$ are about 0.82 and 0.96, respectively. This restricts the scalability of the protocol and implies that its implementation is feasible only with small number of participants and in short-range scenarios.}

Another protocol considered in this paper was three-qubit VBQC. We examined the performance of a test run considering different noise levels at the quantum gates. Our simulation results showed that with our chosen parameters, for a depolarizing probability of $0.05$ at quantum gates, it was not possible to guarantee the security and correctness of this protocol. Furthermore, it can be inferred from the simulation results that if the depolarizing probability at the gate with highest depolarizing noise is at most $0.03$, it is feasible to implement this protocol, although for a limited range of protocol parameters, e.g., $t$ and $w$. To extend the range of feasible protocol parameters, it is necessary to enhance the fidelity of the quantum gates reaching beyond $0.9775$ (corresponding to a depolarizing probability 0f $0.03$). 

Finally, we investigated the performance of quantum digital signature protocol. We evaluated the effect of transmission loss and the error in the measurement setup on the security level of the protocol. Our simulation results showed that the transmission loss adversely affects the probability of repudiation significantly. To compensate for the degrading effect of loss, one can increase the number of transmitted qubits.  

{It is worth noting that in this paper, we have considered three-qubit VBQC protocol (we assumed three qubits at the server), which is a specific case of this protocol. One possible future research direction is to change the number of qubits at the server and examine the impact of this parameter. Moreover, other types of BQC such as multi-server BQC protocols \cite{morimae2013secure,li2014triple,sano2021multi} can be considered. Furthermore, in this work we have assumed equal depolarization probability for all quantum gates to provide benchmarks. It will be interesting to investigate the effect of each individual gate by considering different noise levels for quantum gates. Another possible future research direction is to investigate the robustness of the quantum money protocol to different sources of loss such as transmission loss and quantum memory loss. Finally, one can consider the generalized QDS protocol with more than three parties \cite{amiri2016unconditionally} and evaluate the scalability of this protocol in practical scenarios like metropolitan area networks.}

In summary, in this paper we presented detailed performance analysis of several quantum protocols: quantum money, W-state based anonymous transmission, verifiable blind quantum computation, and quantum digital signature. The simulation results presented in this paper provides a better understanding of advantages and limitations of these protocols and paves the way for efficient design and implementation of these protocols in future quantum networks.

\section*{Data availability}
\urlstyle{rm}
All data generated in this paper can be reproduced by the provided methodology. The code repository for all NetSquid simulations performed in this work is available at: {\url{https://github.com/LiaoChinTe/netsquid-simulation}}.

\section*{Acknowledgements}
We acknowledge support of the European Union’s Horizon 2020 Research and Innovation Program under grant agreement number 820445 (QIA). This work was supported by EPSRC grants EP/N003829/1.

\section*{Appendices}

\section*{Appendix A: Derivation of $Pr(A|B)$ and $Pr(A|C)$}
\label{AT_appendix}
In this appendix, we calculate the two probabilities $Pr(A|B)$ and $Pr(A|C)$. First of all, we note that $Pr(A|B)=1-Pr(A^{\prime}|B)$ and $Pr(A|C)=1-Pr(A^{\prime}|C)$, where $A^{\prime}$ is the event that all measurement outcomes are zero. First we consider the ideal case where there is no noise. It has been shown in \cite{lipinska2018anonymous} that in this case $Pr(A^{\prime}|B)=2/N$. To obtain $Pr(A^{\prime}|C)$, note that the quantum state after loss of one particle is 
\begin{equation}
\rho_{\rm loss}=\frac{N-1}{N}|W\rangle \langle W|_{N-1}+\frac{1}{N} |\vec{0}\rangle \langle \vec{0} |_{N-1} 
\end{equation}
Then, $Pr(A^{\prime}|C)$ is given by 
\begin{eqnarray}
Pr(A^{\prime}|C)={\rm Tr} [\rho_{\rm loss}(\boldsymbol{1}_{SR}\otimes |\vec{0}\rangle \langle \vec{0} |_{N-3})]=
(\frac{N-1}{N})(\frac{2}{N-1})+\frac{1}{N}=\frac{3}{N}\quad \quad \quad \quad \quad \quad \quad
\end{eqnarray}
In the above equation, $\boldsymbol{1}$ denotes identity matrix.

Next, we consider the case where generation and distribution of W state is not ideal. We assume dephasing noise model with the noise parameter $v$ as follows:
\begin{equation}
\Lambda(\rho)=v\rho+(1-v) \sigma_z \rho \sigma_z,
\end{equation}
where $\sigma_z$ is the Pauli Z matrix. The probability $Pr(A^{\prime}|B)$ can be expressed as
\begin{equation}
Pr(A^{\prime}|B)={\rm Tr}[(\Lambda ^{\otimes N} |W\rangle \langle W|_N)(\boldsymbol{1}_{SR}\otimes |\vec{0}\rangle \langle \vec{0} |_{N-2})],
\end{equation}
We note that the term $\boldsymbol{1}_{SR}\otimes |\vec{0}\rangle \langle \vec{0} |_{N-2}$ is a diagonal matrix. Since dephasing noise only affects the non-diagonal elements of the state $|W\rangle \langle W|_N$, we can simply conclude that
\begin{eqnarray}
{\rm Tr}[(\Lambda ^{\otimes N} |W\rangle \langle W|_N)(\boldsymbol{1}_{SR}\otimes |\vec{0}\rangle \langle \vec{0} |_{N-2})]={\rm Tr}[|W\rangle \langle W|_N(\boldsymbol{1}_{SR}\otimes |\vec{0}\rangle \langle \vec{0} |_{N-2})]=\frac{2}{N}.
\end{eqnarray}
From the above equation, we can simply conclude that $Pr(A^{\prime}|C)$ also does not change compared to the noiseless case.

\section*{Appendix B: VBQC protocol}
\label{APP_VBQC}
In this appendix, the steps of the VBQC protocol proposed in \cite{kashefi2020vbqc} are outlined. We assume that the server has three qubits. This protocol consists of $N$ runs, where $d$ runs are computation runs and $t=N-d$ runs are test runs. The client chooses uniformly at random the test runs. In what follows, ${M}^{\alpha}$ denotes a measurement in basis $|\pm _{\alpha}\rangle=(|0\rangle \pm e^{i\alpha}|1\rangle)/\sqrt{2}$. The outcome will be zero for the projector $|+_{\alpha}\rangle\langle +_{\alpha}|$ and 1 for $|-_{\alpha}\rangle\langle -_{\alpha}|$. The inputs of the protocol are as follows:

a) $x \in \{0,1\}$

b) $\phi_1$, $\phi_2$, $\phi_3$ from the set $C=\{k \pi/8\}, k \in [0,7]$

In each test run, the following steps are performed:

 1) Client and server establish three entangled links between them. We denote server's qubits by $q_1$, $q_3$ and $q_5$. Client's qubits entangled with $q_1$, $q_3$ and $q_5$ are denoted by $q_2$, $q_4$ and $q_6$, respectively. 
 
 2) Server applies CZ on $q_1$ and $q_3$. Then, server applies CZ on $q_3$ and $q_5$.
 
 3) Client randomly chooses $u \in \{1,2\}$. 
 
 4) Client randomly chooses  $\theta_1$, $\theta_2$, and $\theta_3$ from the set $C$.
 
 5) if $u=1$, client applies ${M}^{-\theta_1}$ on $q_2$, with the result denoted by $g_1$. The client applies ${M}^{-\theta_3}$ on $q_6$, with the result denoted by $g_3$. The client measures $q_4$ in the standard basis, with the result denoted by $d_2$. 
 
 6) If $u=2$, client applies ${M}^{-\theta_2}$ on $q_4$, with the result denoted by $g_2$. Client measures $q_2$ and $q_6$ in the standard basis, with the results denoted by $d_1$ and $d_3$, respectively.
 
 7) Client randomly chooses $r_1$, $r_2$ and $r_3$ from the set $\{0,1\}$.
 
 8) If $u=1$, client assigns $\delta_1=\theta_1+(r_1+d_2+g_1)\pi$. Otherwise, client randomly chooses $\delta_1$ from the set $C$. Client sends $\delta_1$ to server. Server applies ${M}^{\delta_1}$ on $q_1$ and sends the result, denoted by $b_1$, to client. 
 
 9) If $u=1$, client randomly chooses $\delta_2$ from the set $C$. Otherwise, client assigns $\delta_2=\theta_2+(r_2+d_1+d_3+g_2)\pi$. Client sends $\delta_2$ to the server. The server applies ${M}^{\delta_2}$ on $q_3$ and send the result, denoted by $b_2$, to client.
 
 10) If $u=1$, client assigns $\delta_3=\theta_3+(r_3+d_2+g_3)\pi$. Otherwise, client randomly chooses $\delta_1$ from the set $C$. Client sends $\delta_1$ to server. Server applies ${M}^{\delta_3}$ on $q_5$ and sends the result, denoted by $b_3$, to client.  
 
 11) If $u=1$, client verifies the test round if $r_1=b_1$ and $r_3=b_3$. Otherwise, client verifies the test round if $r_2=b_2$. 
 
In each computation run the following steps are performed:

1) Client and server establish three entangled links between them. We denote server's qubits by $q_1$, $q_3$ and $q_5$. Client's qubits entangled with $q_1$, $q_3$ and $q_5$ are denoted by $q_2$, $q_4$ and $q_6$, respectively. 
 
 2) Server applies CZ on $q_1$ and $q_3$. Then, Client applies CZ on $q_3$ and $q_5$. 
 
 3) Client randomly chooses  $\theta_1$, $\theta_2$, and $\theta_3$ from the set $C$. 
 
 4) Client applies ${M}^{-\theta_1}$, ${M}^{-\theta_2}$, and ${M}^{-\theta_3}$ to $q_2$, $q_4$ and $q_6$, respectively, with the results assigned to $g_1$, $g_2$ and $g_3$, respectively. 
 
 5) Client randomly chooses $r_1$, $r_2$ and $r_3$. 
 
 6) Client sends $\delta_1=\phi_1+\theta_1+(x+r_1+g_1)\pi$ to server. Server applies ${M}^{\delta_1}$ to $q_1$ and sends the result, $b_1$ to client. 
 
 7) Client sends $\delta_2=(-1)^{b_1+r_1}\phi_2+\theta_2+(r_2+g_2)\pi$ to server. Server applies ${M}^{\delta_2}$ to $q_3$ and sends the result, $b_2$ to client.
 
 8) Client sends $\delta_3=(-1)^{b_2+r_2}\phi_3+\theta_3+(b_1+r_1+r_3+g_3)\pi$ to server. Server applies ${M}^{\delta_3}$ to $q_5$ and sends the result, $b_3$ to client. 
 
 9) Client considers $b_3\oplus r_3$ as the output. 
 
 If the number of failed test runs is larger than a threshold the protocol aborts.
\section*{Appendix C: QDS protocol}
\label{QDS_appendix}
In this appendix, the QDS protocol proposed in \cite{amiri2016secure} is explained in details. This protocol has three parties, namely, Alice, Bob, and Charlie. Alice sends the signed message to Bob. Bob authenticates the message and forwards it to charlie. The protocol is consisted of two parts: distribution stage, and messaging stage. 

In the distribution stage, for each possible message $m=0$ or $m=1$, key generation protocol (KGP) is performed by Alice-Bob and Alice-Charlie separately. The number of transmitted qubits in $KGP$ is denoted by $N$. In \cite{amiri2016secure}, weak coherent states with decoy states technique are used in the KGP. Here, we assume KGP is performed using single photon states. To perform KGP between Alice and Bob, Bob randomly chooses from the four states $\left |0_Z\right \rangle$,$\left |1_Z\right \rangle$ ($Z$ basis), and $\left |0_X\right \rangle=1/\sqrt(2)(\left |0_Z\right \rangle+\left |1_Z\right \rangle)$,  $\left |1_X\right \rangle=1/\sqrt(2)(\left |0_Z\right \rangle-\left |1_Z\right \rangle)$ ($X$ basis) and sends the quantum states to Alice. The $X$ and $Z$ bases are selected with probabilities $P_X \geq 0.5$ and $P_Z=1-P_X$, respectively. Alice randomly chooses her measurement basis $X$ or $Z$ with probabilities $P_X$ and $P_Z$, respectively. Then, Alice and Bob perform key sifting to obtain two strings with matched bases. Alice and Bob's key strings are denoted by $A^{B}_m$ and $K^{B}_m=\{X^{B}_m,Z^{B}_m\}$, respectively.
Here, $X^{B}_m$ and $Z^{B}_m$ denote Bob's key in $X$ and $Z$ bases respectively. Similarly, Alice and Charlie perform KGP and generate bit strings $A^{C}_m$ and $K^{C}_m=\{X^{C}_m,Z^{C}_m\}$.

Next, Bob randomly selects a fraction $r$ of $X^{B}_{m}$, denoted by $V^{B}_m$ to estimate the error with Alice. The estimated error rate is denoted by $e^{BA}_X$. Then, he randomly chooses half of the remaining bits and sends them to Charlie. The forwarded and left strings are denoted by $X^{B,\rm forward}_m$ and $X^{B,\rm keep}_m$, respectively. Similarly, Charlie does the same procedure, and obtains $e^{CA}_X$, $X^{C,\rm forward}_m$ and $X^{C,\rm keep}_m$. At the end of the distribution stage, Bob and Charlie have the strings $R^{B}_m=\{X^{B,\rm keep}_m$,$X^{C,\rm forward}_m\}$ and $R^{C}_m=\{X^{C,\rm keep}_m$,$X^{B,\rm forward}_m\}$, respectively. Moreover, Alice has two strings ${\acute{A}}^{B}_m$ and ${\acute{A}}^{C}_m$. The length of these four strings is represented by $l$. 

In the messaging stage, Alice sends $(m,S_m)$ to Bob, where $S_{m}=\{{\acute{A}}^{B}_m,{\acute{A}}^{C}_m \}$. Bob checks the mismatches between $R^{B}_m$ and $S_m$. He will accept the message and forward it to Charlie if there are fewer than $s_a l$ mismatches in both halves of his string, where $0<s_a<0.5$ is a threshold. Similarly, Charlie will accept the message if there are fewer than $s_v l$ mismatches in both halves of his key, where $0<s_v<s_a<0.5$.

In \cite{amiri2016secure}, the thresholds $s_a$ and $s_a$ are chosen to be 

\begin{eqnarray}
s_a=\frac{P_E+2 e^U_X}{3}\\ \nonumber
s_v=\frac{2P_E+e^U_X}{3}.
\end{eqnarray}
In the above equations, $e^U_X=\max \{e^{U}_{X,B},e^{U}_{X,C}\}$, where $e^{U}_{X,B}$ and $e^{U}_{X,C}$ are upper bounds on error rate in X basis. The parameter $e^{U}_{X,B}$ and $e^{U}_{X,C}$ can be obtained using $e^{BA}_X$ and $e^{CA}_X$, respectively. To this aim, Serfling inequality \cite{serfling1974probability} is applied to bound the actual error rate. The parameter $P_E$ is given by 
\begin{equation}
h(P_E)=1-h(\phi ^U_X),
\end{equation}
where $h(.)$ is binary entropy function, and $\phi ^U_X$ is the upper bound on the phase error rate in the $X$ basis. This parameter is obtained using the estimated bit error rate in the $Z$ basis, as in \cite{zhang2017improved}.

The security of the protocol is characterized by $P_{\rm abort}$, $P_{\rm for}$, $P_{\rm rep}$. The parameter $P_{\rm abort}$ is given by
\begin{equation}
P_{\rm abort}=2 \epsilon_{PE},
\label{p_abort}
\end{equation}
where $\epsilon_{PE}$ is the failure probability in calculating the upper bounds for error rates. $P_{\rm rep}$ can be expressed as
\begin{equation}
P_{\rm rep}=2 e^{-0.5l(s_a-s_v)^2}.
\end{equation}
Finally, the parameter $P_{\rm for}$ is given by
\begin{equation}
P_{\rm for}=a+\epsilon_F+8 \epsilon_{PE},
\end{equation}
where
\begin{equation}
\epsilon_F=\frac{1}{a}(\epsilon+2^{(-l(1-h(\phi ^U_X)-h(s_v))}).
\end{equation}
In the above equation, $a$ and $\epsilon$ are small constants. 

\section*{Appendix D: NetSquid implementation}
\label{NetSquidImplementation}
In this appendix, we provide some details on NetSuid simulations. The code repository for all NetSquid simulations performed in this work can be found in \cite{netsquidGithub}. A README file is included for each protocol, which provides a general description of the protocol, protocol steps, protocol parameters, and other necessary details. 

\bibliography{Reference_SR}

\end{document}